\documentclass[preprint,showpacs,showkeys]{revtex4}
\usepackage{graphics,graphicx}
\usepackage{amssymb}
\usepackage{amsmath}
\usepackage{bm}
\begin{document}
\setcounter{page}{0}
\title[]{Magnetization plateaus in generalized Shastry-Sutherland models}
\author{Keola \surname{Wierschem}}
\email{keola@ntu.edu.sg}
\author{Pinaki \surname{Sengupta}}
\email{psengupta@ntu.edu.sg}
\affiliation{Division of Physics and Applied Physics, Nanyang Technological University, Singapore}

\date{Received 5 June 2012}

\begin{abstract}

We study an anisotropic Heisenberg antiferromagnet with ferromagnetic transverse spin exchange using exact quantum Monte Carlo methods. Such a model is relevant to a class of rare earth tetraboride materials that display a range of magnetization plateaus under applied magnetic field. The layered arrangement of magnetic ions in these materials is topologically equivalent to the Shastry-Sutherland lattice. In this frustrated geometry, we study the interplay of next-nearest neighbor interactions in stabilizing a plateau at half the saturation magnetization (or 1/2 plateau). We also show hysteresis-like behavior at the onset of the 1/3 plateau.

\end{abstract}



\maketitle

\section{INTRODUCTION}

For many years, the Shastry-Sutherland lattice (SSL) remained a purely theoretical model of frustrated magnetism~\cite{shastry1981}. Then, a little over a decade ago, magnetization plateaus were discovered in the SSL material SrCu$_2$(BO$_3$)$_2$ under high magnetic fields~\cite{kageyama1999}. More recently, the rare earth tetraborides were found to form an entire class of SSL compounds displaying a range of magnetization plateaus~\cite{yoshii2006}. Thus, the SSL has been realized in magnetic materials and is accompanied by interesting physical behavior.

The rare earth tetraborides occur as tetragonal crystals with an SSL arrangement of rare earth ions in the crystal plane and alternating dimer bonds of roughly equal strength to those of the square lattice bonds~\cite{yoshii2006}. The rare earth tetraborides are very interesting because not only do they display plateau structure, but they also have low saturation fields that allows for the complete experimental characterization of magnetic behavior.

From the theoretical point of view, the relatively large single-ion anisotropy of these materials, coupled with a large total angular momentum, allows for the formation of an effective spin-1/2 XXZ model with strictly ferromagnetic transverse spin exchange. Such a model is free from the sign problem normally associated with frustrated spin systems, and can thus be directly simulated using very precise quantum Monte Carlo (QMC) techniques. Originally a classical Ising model was employed to understand the origin of the magnetic plateaus~\cite{siemensmeyer2008}, but it was quickly realized that the 1/2 plateau seen for example in TmB$_4$ cannot be reproduced on the SSL. Such a phase only stabilizes for the XXZ model near the Ising limit~\cite{meng2008}. Even so, the plateau sequence observed in TmB$_{4}$ has no 1/3 plateau, and to replicate this with the effective spin-1/2 model it was necessary to add extended bonds to the SSL~\cite{suzuki2009, suzuki2010}.

\begin{figure}
\includegraphics[clip,trim=0cm 10cm 0cm 10cm,width=0.75\linewidth]{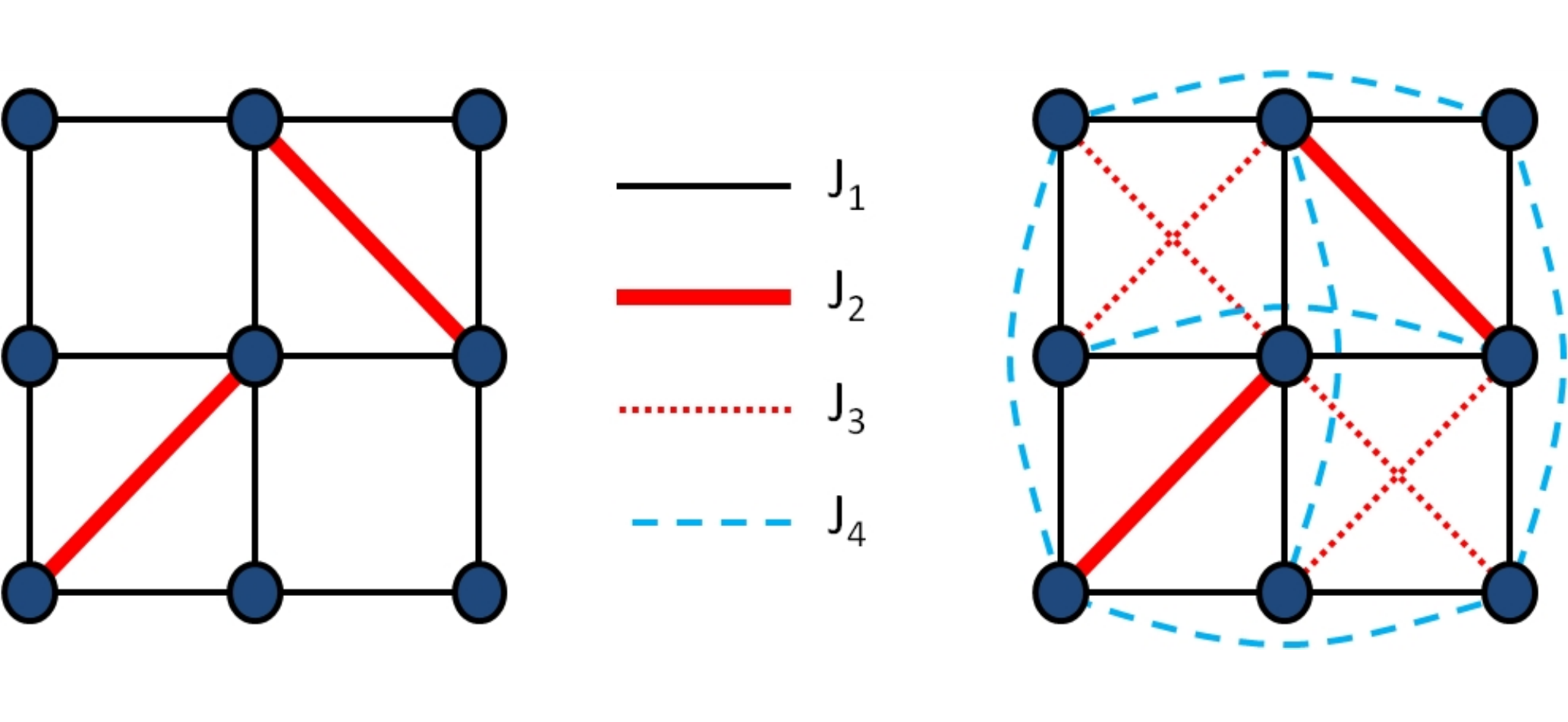}
\caption{(Color online) Illustration of bonds in the Shastry-Sutherland lattice (left) and the extended Shastry-Sutherland lattice (right).}\label{fig-lattice}
\end{figure}

The model Hamiltonian of the anisotropic Heisenberg antiferromagnet with ferromagnetic transverse exchange is
\begin{equation}
\label{eq-model}
H=\sum_{\alpha}\sum_{\left<ij\right>_{\alpha}}\left[-\left|\Delta J_{\alpha}\right|\left(S_i^xS_j^x+S_i^yS_j^y\right)+J_{\alpha}S_i^zS_j^z\right]-h\sum_{i}S_{i}^{z}
\end{equation}
where the transverse exchange coupling $-|\Delta J_{\alpha}|$ is always ferromagnetic and we sum over all bond types $\alpha$ and their corresponding site indices $<ij>_{\alpha}$. For the original Shastry-Sutherland lattice, there are two species of bonds. First, for $\alpha=1$, we have the typical bonds of a square lattice. Second are the $\alpha=2$ bonds that form alternating dimers and introduce frustration. Continuing on to an extended Shastry-Sutherland lattice, we define $\alpha=3$ bonds along the diagonals of plaquettes not containing an $\alpha=2$ bond, while $\alpha=4$ bonds are a simple extension of the $\alpha=1$ bonds (see Fig.~\ref{fig-lattice} for an illustration).

In this manuscript, we present the results of a QMC investigation of the above Hamiltonian in the parameter regime of relevance to the rare earth tetraborides, namely $J_{1}=J_{2}$ and $\Delta\ll1$. Using the stochastic series expansion QMC method~\cite{sandvik1999}, we confirm the findings of Refs.~\cite{suzuki2009,suzuki2010} that the addition of a ferromagnetic $J_{4}$ coupling tends to stabilize the 1/2 plateau and eliminate the 1/3 plateau. Further, we show that this effect is largely independent of the presence of an antiferromagnetic $J_{3}$ coupling. We also study the history-dependent behavior of the magnetization response for increasing and decreasing magnetic field strengths.

\section{Results and discussion}

\begin{figure}
\includegraphics[clip,trim=0cm 1cm 0cm 1cm,width=0.75\linewidth]{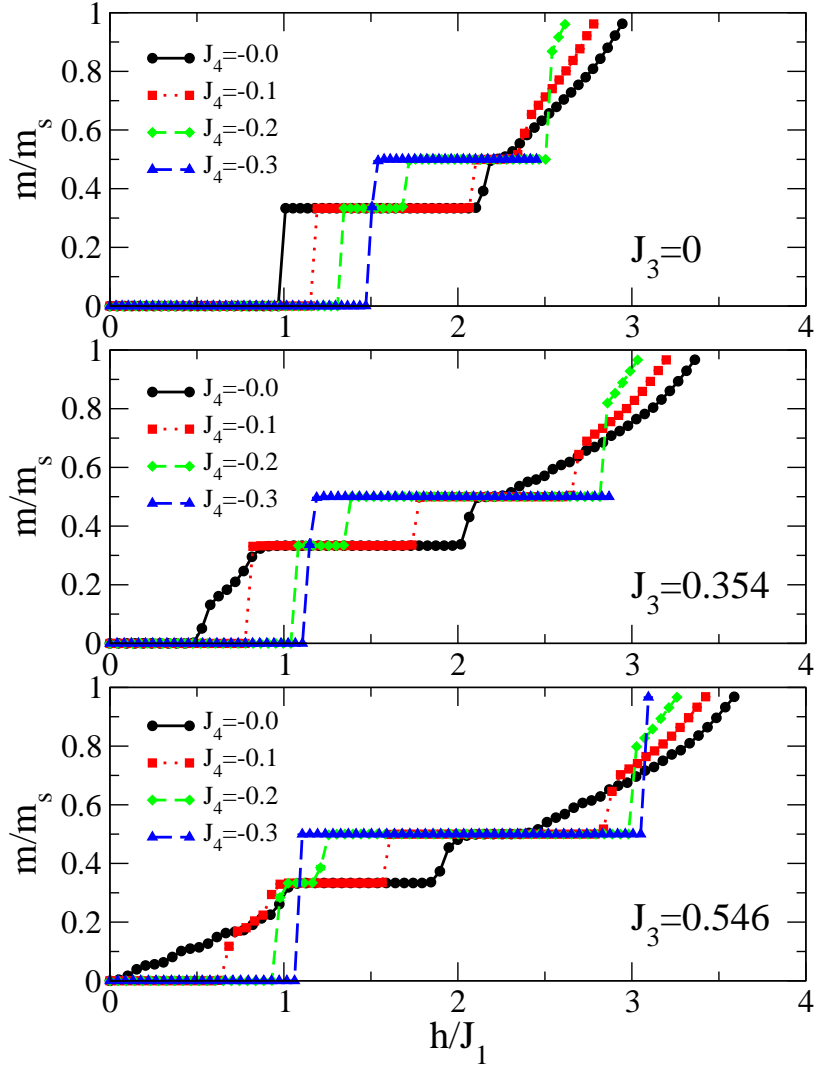}
\caption{(Color online) Field dependence of the uniform magnetization as the magnitude of $J_4$ is increased for three $J_3$ values. Results are from a simulation on a $6\times6$ lattice with $\Delta=0.178$ and $J_{1}=J_{2}$.}\label{fig-matas}
\end{figure}

The rare earth tetraborides are {\em metallic}, and thus the RKKY interaction can be expected to provide the spin-exchange coupling between localized lattice spins. This provides a motivation for including interactions between extended neighbors, as well as an expectation of ferromagnetic interactions for distant neighbors.

In Fig.~\ref{fig-matas} we present our results for the magnetization behavior for a variety of parameter values. In each panel, data from several $J_{4}$ values is presented side-by-side, with each panel using data from a different $J_{3}$ value. Several trends are discernible in this figure. The width of the 1/2 plateau increases with increasing $J_{3}$. The critical magnetic field at which N{\'e}el order disappears shifts to lower field values, or disappears entirely. However, it is noteworthy that the critical value $J_{4}^{*}$ at which the 1/3 plateau disappears is in the range $-0.3<J_{4}^{*}<-0.2$ for all values of $J_{3}$. Thus, the value of $J_{4}$ needed to replicate the magnetization curve of TmB$_{4}$ is somewhat independent of the $J_{3}$ parameter. A finer grid of  $J_{4}$ values is needed to determine what, if any, dependence there is.

\begin{figure}
\includegraphics[clip,trim=0cm 1cm 0cm 1cm,width=0.75\linewidth]{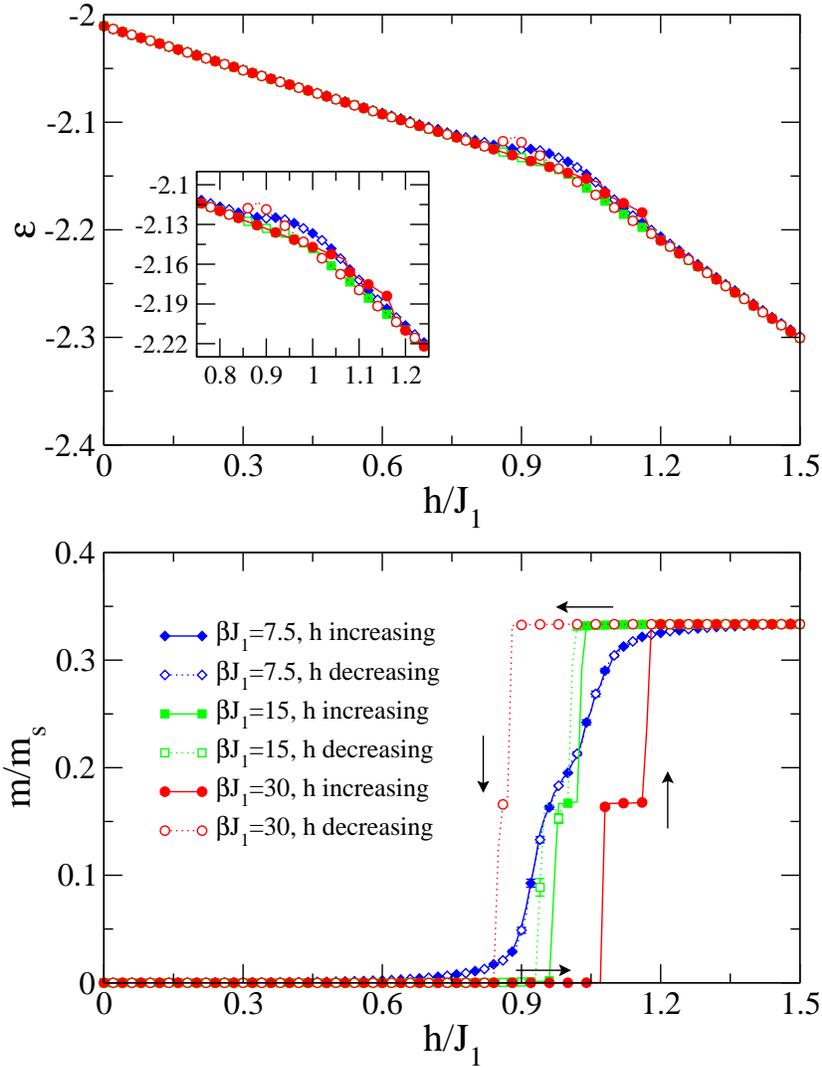}
\caption{(Color online) Energy per particle (a) and magnetization fraction (b) of a $12\times12$ lattice for increasing  and decreasing magnetic field values with $\Delta=0.2$, $J_{1}=J_{2}$ and $J_{3}=J_{4}=0$. The inverse temperature legend in (b) also applies to (a) and its inset, which shows a close-up of the transition region. Arrows drawn in (b) clarify the direction of the magnetization curve for increasing and decreasing magnetic field. Note that not all data points are represented by symbols (though the connecting lines pass through them).}\label{fig-ramp}
\end{figure}

To investigate the magnetic hysteresis of the model Hamiltonian in Eq.~\ref{eq-model}, we have performed careful ramp-up and ramp-down studies where the magnetic field is slowly increased (ramp-up) or decreased (ramp-down). In between these changes in magnetic field, many QMC iterations are performed to reach a quasi-equilibrium before continuing at the next magnetic field value. As the starting configuration at each magnetic field value is taken from the preceding point in the ramp-up or ramp-down procedure, any history-dependence (i.e. hysteresis) should be captured in a comparison of the ramp-up and ramp-down processes.

As seen in Fig.~\ref{fig-ramp}, the energy per particle reaches the low-temperature limit by $\beta J=15$. At this inverse temperature and below, the ramp-up and ramp-down energies are the same to within our statistical uncertainty. However, at $\beta J=30$ a clear hysteresis is seen near the critical field $h_c$ where a cusp exists in the energy curve. The crossing point of the ramp-up and ramp-down energy curves gives a nice estimate of $h_c$. Likewise, the ramp-up and ramp-down behavior for the magnetization curve is history-independent at inverse temperatures below $\beta J=15$, while hysteresis sets in at $\beta J=30$. Right at $\beta J=15$, critical fluctuations of the magnetization are seen near $h_c$, indicative of a peak in the static susceptibility.

\section{CONCLUSIONS}

We have presented the results of a quantum Monte Carlo investigation of the Ising-like XXZ Heisenberg model in the parameter regime of relevance to the rare earth tetraborides, namely $J_{1}=J_{2}$ and $\Delta\ll1$. There are a number of conclusions to be made from this study. First, we confirm the findings of Refs.~\cite{suzuki2009,suzuki2010} that the addition of a ferromagnetic $J_{4}$ coupling tends to stabilize the 1/2 plateau and eliminate the 1/3 plateau. Second, we show that this effect is largely independent of the presence of an antiferromagnetic $J_{3}$ coupling. Last, we have shown history-dependent behavior of the magnetization response for increasing and decreasing magnetic field strengths.

\end{document}